\documentclass[]{aa}
\usepackage{graphicx}
\usepackage{txfonts}
\begin{document}
   \title{Slow evolution of elliptical galaxies induced by dynamical
    friction}

   \subtitle{II. Non-adiabatic effects}

   \author{S.E. Arena,
          \inst{1}
          G. Bertin,
          \inst{1}
          T. Liseikina,
          \inst{2}
          \and
          F. Pegoraro,
          \inst{3}
          }

   \offprints{S.E. Arena}

   \institute{Universit\`{a} degli Studi di Milano, Dipartimento di Fisica, via
             Celoria 16, I-20133 Milano, Italy\\
         \email{Serena.Arena@unimi.it}\\
             \email{Giuseppe.Bertin@unimi.it}
         \and
             Ruhr-Universitaet, Bochum, Germany\\
             \email{tatianavl@ngs.ru}
         \and
             Universit\`{a} degli Studi di Pisa, Dipartimento di Fisica, largo Pontecorvo 3, 56100 Pisa, Italy\\
             \email{pegoraro@df.unipi.it}
             }

   \date{Received 21 January 2006 / Accepted 14 March 2006}

  \abstract
   {Many astrophysical problems, ranging from structure formation in
     cosmology to dynamics of elliptical galaxies, refer to
     slow processes of evolution of essentially collisionless self-gravitating systems.
     In order to determine the relevant quasi-equilibrium configuration at time $t$ from
     given initial conditions, it is often argued that such slow evolution may be
     approximated in terms of adiabatic evolution, for the calculation of which
     efficient semi--analytical techniques are available.}
   {Here we focus on the slow
     process of evolution, induced by dynamical friction of a host stellar system
     on a minority component of ``satellites", that we have investigated in a
     previous paper, to determine to what extent an adiabatic description might be applied. }
   {The study is realized by comparing directly N--body
     simulations of the stellar system evolution (in two significantly different models) from initial to final conditions in a
     controlled numerical environment.}
   {We demonstrate that for the examined process the
     adiabatic description is going to provide incorrect answers, not only quantitatively,
     but also qualitatively. The two classes of models considered exhibit generally similar
     trends in evolution, with one exception noted in relation to the
     evolution of the \emph{total} density profile.}
   {This simple conclusion should be
     taken as a warning against the indiscriminate use of adiabatic
     growth prescriptions in studies of structure of galaxies.}


   \titlerunning{Non-adiabatic slow evolution of stellar systems}
   \authorrunning{S.E. Arena et al.}
   \maketitle
%

\section{Introduction}

For the study of essentially collisionless systems the paradigm of
adiabatic growth has received significant attention in
astrophysics, especially in relation to two important contexts.
Within cosmological scenarios of structure formation, it has been
proposed as a natural framework to calculate the effects of infall
of baryonic matter into the potential wells dominated by dark
matter halos (see Blumenthal et al. 1986; Mo, Mao \& White 1988;
Kochanek \& White 2001; Gnedin et al. 2004). In addition, it has
been used as a tool to model stellar dynamical cusps in the
vicinity of massive black holes at the center of galaxies (Young
1980; Goodman \& Binney 1984; Cipollina \& Bertin 1994; Quinlan,
Hernquist \& Sigurdsson 1995; Merritt \& Quinlan 1998; van der
Marel 1999). Either by semi-analytical techniques or by means of
suitable simulations, it has thus been shown that such adiabatic
growth (usually justified by recalling that dissipative matter is
able to accrete slowly towards the center) leads to a steepening
of the underlying potential well.

Recently, in contrast with the above picture, several
investigations have noted that processes usually interpreted in
terms of dynamical friction, such as the capture of satellites and
merging, tend to lead to systems with softer density profiles
(El-Zant et al. 2001; Bertin, Liseikina \& Pegoraro 2003,
hereafter Paper I; El-Zant et al. 2004; Ma \& Boylan-Kolchin 2004;
Nipoti et al. 2004), with a significant impact on the current
debate about the observational counterparts to the universal halo
density profiles found in cosmological simulations (Navarro, Frenk
\& White 1996; Moore et al. 1998; Ghigna et al. 2000; Navarro et
al. 2004).

In Paper I we set up a numerical laboratory for the study of
dynamical friction by simulating the slow evolution of a stellar
system as a result of the interaction with a system of satellites,
with a distribution that guarantees that approximate spherical
symmetry is maintained. Such quasi-spherical model (inspired by
the observation of globular cluster systems in some elliptical
galaxies, but not necessarily meant to reproduce those conditions)
has the advantage of being associated with a very smooth
evolution, as desired if we wish to identify the small effects
that are characteristic of dynamical friction; in this sense, it
extends and improves earlier models, based on the capture of a
single satellite, used to study the mechanism of dynamical
friction (see Bontekoe \& van Albada 1987; Zaritsky \& White 1988;
Bontekoe 1988; and many following papers).

The process of dynamical friction is not expected to follow the
adiabatic picture, because the ``microscopic" mechanisms at its
basis (scattering of orbits in a discrete picture or wave-particle
resonance in a continuum Vlasov description) are inherently
non-adiabatic. In this paper we clarify this point by analyzing a
set of controlled numerical experiments aimed at pointing out the
differences between slow non-adiabatic and slow adiabatic
evolution. For the general concepts and definitions relevant to
the description of adiabatic processes in classical mechanics, the
reader is referred to classical books such as Landau \& Lifchitz
(1969) (see Chapter 49), Goldstein (1980) (see Chapter 11-7), or
advanced monographs such as Northrop (1963).

In Sect.~2 we describe the planned experiments and diagnostics,
both at macroscopic and at microscopic level. In Sect.~3 we
present the models and the code adopted in the numerical
simulations. In Sect.~4 we describe the results, demonstrate the
significance of non-adiabatic effects in phase space, and show
that the evolution of the underlying distribution function of a
stellar system can be qualitatively different in non-adiabatic
models from that found in the adiabatic case. In Sect.~5 we draw
the main conclusions. In the Appendix we provide a theoretical
framework to the adopted diagnostics of ``scatter plots" by
introducing a suitable description of the evolution of the
relevant distribution function in phase space.


\section{Two experiments of slow evolution within quasi--spherical symmetry}

No dynamical evolution is expected for a stellar system in a
(stable) collisionless equilibrium. But in the presence of some
collisionality  such steady state can be broken. This is the case
for a stellar system that drags in, by dynamical friction, a heavy
object or a system of heavy objects toward its center; while these
``satellites" are dragged inwards, the reaction to the effects of
dynamical friction makes the stellar system evolve (see Paper I),
although very slowly. The question that we wish to clarify by
means of a set of dedicated experiments is to what extent such
slow evolution differs from a process in which a secular shift of
matter toward the center of the stellar system occurs
adiabatically.

To study this problem, following Paper I we refer to a spherically
symmetric stellar system hosting a shell of total mass $M_{shell}$
fragmented into $N_{shell}$ heavy objects. In the continuum limit,
of $N_{shell} \rightarrow \infty$ at fixed shell mass, the system
would be characterized by spherical symmetry; for finite values of
$N_{shell}$, the system is only approximately spherically
symmetric. The symmetry of the system and the focus on a shell are
chosen so as to better identify the small effects of slow
evolution in the mechanisms that we wish to investigate.

In the following subsections we describe two experiments designed
to study the differences between adiabatic and non-adiabatic
processes.

\subsection{A shell distribution of satellites dragged in by dynamical
friction (a non-adiabatic process)} \label{caseA}

The first experiment, identified as case $A$, is the same as for
runs $BS_1$ and $BS_2$ described in Paper I, that is a case when
the evolution of the host galaxy proceeds as a result of the
interaction with a discrete set of heavy objects falling in
because of dynamical friction; while such a shell of satellites
falls slowly toward the center, the galaxy rearranges its stellar
orbits and therefore changes its distribution function.

The galaxy is taken to be described by a suitable equilibrium
distribution function $f$. An initially smooth, spherically
symmetric shell density distribution of matter, associated with
the potential $\Phi_{shell}(r)$, is then added to the galaxy. We
recall that, as described in Paper I (Sect.~3.2.2), the initial
self-consistent galaxy potential $\Phi_G(r)$, in the presence of
the additional potential $\Phi_{shell}(r)$, is different from that
associated with $f$ alone (i.e., in the absence of the shell), and
has to be calculated separately to ensure that the initial state
is indeed very close to dynamical equilibrium. The shell is then
fragmented into $N_{shell}$ fragments (called also satellites)
with initial velocities that are appropriate for test particles on
circular orbits in the combined potential generated by the galaxy
and the initially smooth shell ($\Phi_G(r)+\Phi_{shell}(r)$).

In the experiment, the satellites interact directly with the
simulation particles that represent the collisionless host galaxy,
and thus experience dynamical friction, spiraling in toward the
center on a time-scale $\tau_{fr}$ much longer than the dynamical
time $t_d$. The evolution of the whole system is expected to be
\emph{slow, but non--adiabatic}.

\subsection{A similar shell distribution of matter moved in adiabatically}

The second experiment, identified as case $B$, considers the
evolution of the same collisionless host galaxy considered in case
A, but in the presence of an \emph{external} collisionless shell
of matter, which is moved (arbitrarily) toward the center of the
galaxy mimicking the slow sinking of the fragmented shell of case
A with the same time--scale $\tau_{fr} \gg t_d$. In this case the
dynamical friction process is not in action and the galaxy evolves
because the external field experienced is (slowly) time-dependent.

In contrast with case $A$, here the shell is always kept smooth,
spherically symmetric, not fragmented, and made to act on the
galaxy uniquely as a slowly varying ``external field". The
evolution of the galaxy is thus expected to be, by construction,
\emph{slow and adiabatic}.

\subsection{Control cases}

We will also perform two additional experiments devised to test
the dynamical evolution of the system when the mechanisms
described above are not applicable.

The first control experiment, identified as case $C$, is a
simulation that follows the prescription of case $B$, but with a
much shorter time--scale $\tau_{fast}$, so as to break, on
purpose, the condition of adiabatic evolution. This experiment is
meant to better identify those properties that are characteristic
of adiabatic evolution with respect to those associated with the
driving by a general external forcing.

The second control experiment, case $D$, is a simulation to test
the quality of the initial equilibrium in each of the above three
cases $A$, $B$, and $C$. We consider two possibilities. One, in
which the shell is treated as smooth, spherically symmetric, and
collisionless, is better suited as a control test for case $A$;
the shell is here represented by a large number of simulation
particles placed on circular orbits but otherwise
indistinguishable from the simulation particles used to represent
the host galaxy. The other, in which the shell only acts as an
``external field", is the more natural control test for cases $B$
and $C$. Case $D$, in addition to testing that the adopted initial
conditions for the model of the galaxy and of the shell are
correct, checks the accuracy of our simulations over long
time-scales.

\subsection{Diagnostics} \label{diag}

To test the similarities and the differences in these cases, we
resort to the following diagnostics. To study the problem at the
macroscopic level, we show for the various cases the evolution of
the galaxy density profile and of the global pressure anisotropy,
defined as $2 K_r / K_T$, where $K_r$ and $K_T$ are the total
kinetic energy of the galaxy in the radial and tangential
direction respectively. At the microscopic level, to study the
detailed conservations in phase--space, we produce ``scatter
plots" (final vs. initial quantities) for single--particle energy
$E$, radial action $I$, and angular momentum $J$.

Ideally, in simulations belonging to case $A$, the three
quantities, $E$, $I$, and $J$, are not conserved; in case $B$, the
single--particle energy $E$ is not conserved, but $I$ and $J$ are;
in case $C$, only $J$ is conserved. In case $D$, all quantities,
$E$, $I$, and $J$, are conserved, and the relevant scatter plots
thus define the limits of the numerical simulations (see Tab.
\ref{Tab01}).

The use of scatter plots raises interesting and important
theoretical issues. Consider the scatter plot for the
single-particle energy, which we already presented in the bottom
right frame of Fig.~8 in Paper I and we will give in the upper
frames of Fig.~\ref{Fig03} and Fig.~\ref{Fig04} of the present
paper. This plot correlates the value of the energy of a single
{\it simulation particle} at an initial time $t_{in}$ with that of
the same {\it simulation particle} at a final time $t_{fin}$. In
an ideal code the simulation particles would follow the
characteristics dictated by the collisionless Boltzmann (Vlasov)
equation. Under conditions for which the energy is conserved, any
initial distribution function would originate a distribution of
simulation particles for which the scatter plot is a segment, for
any value of $t_{fin}$; for a finite number of particles, the
points would line up on such segment, covering the range from the
minimum to the maximum value of particle energy, consistent with
the adopted distribution function. Under conditions for which
energy is not conserved, we may expect a distortion of that
segment and, in addition, a {\it broadening} of the segment into a
distribution of points, corresponding to a spread of effects, at
given initial energy, associated with the different values of the
angular momentum. Some conditions that determine the energy
non-conservation may even determine a change in the value of the
energy that depends on {\it phase variables} that define the orbit
of the simulation particles, at given initial values of energy and
angular momentum.

In real simulations, all the effects mentioned above are mixed
with numerical effects, which we may generically call numerical
noise. In other words, undesired numerical noise may be
responsible for distortions of the segment-type plot and, more
naturally, a general broadening with respect to the ideal case.

These considerations prompted us to move beyond the intuitive
stage at which scatter plots are naturally appreciated, to see
whether the information contained there can be set in a more
satisfactory quantitative theoretical framework. An outline of
such theoretical framework is provided in the Appendix.

\begin{table}
\begin{minipage}[t]{\columnwidth}
\caption{Expected conservation of single-particle quantities for
the four cases considered (C: quantity conserved, NC: quantity not conserved).}             
\label{Tab01}      
\centering                          
\renewcommand{\footnoterule}{}  
\begin{tabular}{c | c c c c}        
\hline\hline                 
\textbf{Case} & \textbf{A} & \textbf{B} & \textbf{C} & \textbf{D}\\    
\hline                        
   $E$ & NC & NC & NC & C \\
   $J$ & NC & C  & C  & C \\
   $I$ & NC & C  & NC & C \\
\hline                                   
\end{tabular}
\end{minipage}
\end{table}


\section{The models and the code}

\subsection{The basic equilibrium of the host galaxy}

The basic model for the initial conditions of the host galaxy is
constructed from two different distribution functions. The first
is an $n=3$ polytrope with

\begin{equation}
f (\vec{r},\vec{v}) = A (E_0 - E)^{n-3/2}
\end{equation}

\noindent for $E < E_0$ and zero otherwise. The single particle
total energy is $E = v^2 /2 + \Phi(r)$ and $E_0 = \Phi(R) = - G
(M_G + M_{shell}) / R$, where $M_G$ and $R$ are respectively the
mass and the radius of the galaxy and $M_{shell}$ is the mass of
the shell. This is the galaxy model used in Paper I; it is
convenient and is best suited for comparison with past
investigations of the process of dynamical friction (for example,
it was used by Bontekoe \& van Albada (1987) in their pioneering
work on the subject).

The second is the $f^{(\nu)}$ distribution function

\begin{equation}
f^{(\nu)} = A e^{-[aE +d (J^2/|E|^{3/2})^{\nu/2}]}
\end{equation}

\noindent for $E<0$ and zero otherwise. Here $E$ and $J$ are the
single-star energy and angular momentum. The four constants $A$,
$a$, $d$, and $\nu$ are real and positive: from these one can
identify two dimensional scales and two dimensionless parameters,
such as $\nu$ and the dimensionless central potential $\Psi = -a
\Phi(0)$. In this paper we will focus on the case
$(\nu,\Psi)=(3/4,5)$ (see Trenti \& Bertin (2005) for details
about this distribution function and a discussion of the
properties of the resulting self-consistent models for different
values of $\nu$ and of $\Psi$). The resulting models are more
concentrated than the $n = 3$ polytrope and are characterized by
an anisotropic pressure tensor. In general, these models provide a
more realistic description of bright elliptical galaxies.

\subsection{The modified self--consistent equilibrium in the presence of a shell} \label{SubSec01}

The initial shell is treated as smooth, spherically symmetric, and
static, characterized by the density distribution:

\begin{equation}
\rho_{shell} = \rho_0 \exp{[- 4(r-r_{shell}(0))^2/R_{shell}^2]}
\end{equation}

\noindent for $|r - r_{shell}(0)|\le R_{shell}$ (and vanishing
otherwise); $r_{shell}(0)$ is the initial radial position of the
shell, where its density is $\rho_0$, and $R_{shell}$ is its
half-thickness.

As described in Subsection 3.2.2. of Paper I, in the presence of
an external potential the distribution function of the galaxy
differs, from that in absence of it, in the potential term that
now is the total potential of the system:
$\Phi(r)=\Phi_G(r)+\Phi_{shell}(r)$. As a consequence, it is
necessary to re-compute the equilibrium models of the galaxy by
solving the Poisson equation for the self-consistent $\Phi_G(r)$
in the presence of the external shell density distribution.

\subsection{The code}

The galaxy evolution is described by a collisionless (mean-field)
particle-mesh code, which is an improved version of the original
van Albada (1982) code. The new code (see Trenti et al. (2005) and
Trenti (2005)) is based on an improved Poisson solver scheme, in
which the angular grid is dropped, while the radial grid is
adaptive so that each cell is populated by approximately the same
number of particles.

The galaxy is sampled with $N$ simulation particles, with initial
positions and velocities extracted from the distribution functions
described in subsection \ref{SubSec01} by means of Monte Carlo
methods (see Paper I). The same method is applied to derive the
initial positions of the $N_f$ fragments of the shell, of mass
$M_f = M_{shell}/N_f$. These fragments or satellites are described
by an extended Plummer density profile:

\begin{equation}
\Phi_f(r)=-\frac{GM_f}{(R_f^2+r^2)^{1/2}},
\end{equation}

\noindent where $R_f$ is the radius of each shell fragment. In the
simulations of case $A$ (see subsection \ref{caseA}), the
equations that describe the interactions among galaxy simulation
particles and shell fragments are those described in Paper I. Each
particle of the galaxy feels the action of the whole galaxy via
its mean field and that of each shell fragment directly; each
shell fragment feels the sum of the direct actions from all the
galaxy simulation particles and of the direct actions from the
other shell fragments.

In the simulations of cases $B$ and $C$, there are no discrete
objects (satellites) involved. Therefore, the simulation particles
follow the orbits dictated by the mean field generated by the
galaxy, as in case A, together with the force of the shell which
is treated as an external, spherically symmetric, and smooth
force.

For the diagnostics mentioned in subsection \ref{diag}, the radial
action $I$ of a single simulation particle, at time $t$, is
computed in a semi--analytical way, starting from the values of
$E$ and $J$. The total potential
$\Phi(r)=\Phi_{G}(r)+\Phi_{shell}(r)$, felt by the particle, is
first measured in the simulation at time $t$. Then, the effective
potential $\Phi_{eff}(r;J) = \Phi(r) + J^2/ (2 r^2)$, the radial
velocity $v_r(r;E,J) = \sqrt{2 [E - \Phi_{eff}(r)]}$, and the two
turning points $r_{min}(E,J)$ and $r_{max}(E,J)$ are derived.
Finally, the radial action is found by numerical integration:
$I(E,J) = 2 \int_{r_{min}}^{r_{max}} v_r(r;E,J) dr$.

\subsection{Units}\label{units}

The units adopted are $10$ kpc for length, $10^{11} M_{\odot}$ for
mass, and $10^{8}$ yr for time. Thus, velocities are measured in
units of $97.8$ km/s and the value of the gravitational constant
$G$ is $\approx 4.497$.


\section{Numerical simulations} \label{sim}

The galaxy is sampled with $N = 250000$ simulation particles and
has a mass $M_G = 2$ and a half-mass radius $r_M = 0.3$ (in code
units); simulations with higher values of $N$ have also been run.
We should recall that the issues of the reliability (stability) of
the code and of the adequacy of the number of particles used in
the simulations were already addressed in previous papers. In
particular, in Paper I (see Sect.~5.1 of that paper) we ran a
simulation of a single sinking satellite with $N$ up to $2 \times
10^6$ and found very little change in the observed evolution, down
to the innermost radii reached by the infalling satellite; in
addition (see Sect.~4.4.2 of that paper), we checked, with general
qualitative agreement, the main results obtained earlier by
Bontekoe \& van Albada (1987), who had originally investigated the
process of dynamical friction with only a few thousand simulation
particles. Furthermore, the new version of the code used in this
paper has been extensively tested with many simulations with $N
\geq 8 \times 10^5$, especially designed to check its stability in
reproducing concentrated models over long time scales; the new
version of the code has also been tested against the performance
of high-quality tree codes currently available (Trenti 2005;
Trenti et al. 2005). To be sure, if the process of dynamical
friction depended too sensitively on {\it detailed} resonant
effects, even with these large numbers of simulation particles the
phase space might be insufficiently well sampled.

The mass of the shell is one tenth of that of the galaxy
($M_{shell} = 0.2$). The shell is initially located at
$r_{shell}(0) = r_M$. The radius of each fragment is equal to the
half thickness of the shell, $R_f = R_{shell} = 0.1$, that is one
third of the half-mass radius of the galaxy. These values of
$M_{shell}/M_G$ and of $R_f/r_M$ repeat the conditions of previous
investigations (Bontekoe \& van Albada 1987; Paper I); this choice
helps to better bring out the effects of dynamical friction within
the computational resources used.

With the above choice of $r_M$ and $M_G$, the dynamical time of
the galaxy without the shell, defined as $t_d = GM_G^{5/2} / (- 2
E_{tot})^{3/2}$, is $t_d^{(pol)} = 0.1476$ for the polytropic
model and $t_d^{(\nu)} = 0.1881$ for the $f^{(\nu)}$ model (in
code units). To better illustrate the meaning of $t_d$, we note
that the period of two selected circular orbits, at radii
$1.17~r_M$ and $3.3~r_M$, is $5.4~t_d$ and $14.3~t_d$ for the
polytropic model and $3.0~t_d$ and $12.7~t_d$ for the $f^{(\nu)}$
model.

In all simulations the conservation of the total energy and total
angular momentum is respectively of the order of $10^{-6}$ and
$10^{-5}$ per dynamical time, except for case $D$, for the
polytropic model, in which the conservation is better by one order
of magnitude.

The evolution of the system is followed up to a final time,
$t_{fin} = 90 t_d$ (for the fast case $C$, the evolution is
followed up to $t_{fin} = 2 t_{d} $). The configuration of the
system at such final time is compared with that at $t_{in} =
0.14~t_d$.

The list of experiments is given in Tab.~\ref{Tab02}.

\begin{table}
\begin{minipage}[t]{\columnwidth}
\caption{The main set of numerical simulations; $n=3$ for the
polytropic model and $(\nu,\Psi) = (3/4,5)$ for the $f^{(\nu)}$
model (see text for description); in runs $D1$ and $D3$, $N_f$
represents the number of simulation particles treated as
collisionless.}
\label{Tab02}      
\centering                          
\renewcommand{\footnoterule}{}  
\begin{tabular}{c c c c c}        
\hline\hline                 
\textbf{Case} & \textbf{Process} & \textbf{Run} & \textbf{Shell} &
\textbf{Galaxy} \\
 & \textbf{} &  & \textbf{type} & \textbf{model} \\
\hline                        
   A & Dynamical & A1 & $N_{f}=20$ & Polytrope \\      
     & friction & A2 & $N_{f}=100$  & Polytrope \\
     & & A3 & $N_{f}=20$ & $f^{(\nu)}$ \\
     & & A4 & $N_{f}=100$  & $f^{(\nu)}$ \\
  & & & & \\
   B & External & B1 & External slow  & Polytrope \\
     & adiabatic & B2 & External slow  & $f^{(\nu)}$ \\
  & & & & \\
   C & External & C1 & External fast  & Polytrope  \\
     & non & C2 & External fast  & $f^{(\nu)}$  \\
  & adiabatic& & & \\
   D & Control & D1 & $N_{f}=25000$ & Polytrope  \\
     & & D2 & External static & Polytrope  \\
     & & D3 & $N_{f}=25000$ & $f^{(\nu)}$  \\
     & & D4 & External static & $f^{(\nu)}$  \\
\hline                                   
\end{tabular}
\end{minipage}
\end{table}

\subsection{Case A: Fall of a shell of satellites by dynamical friction}

We have performed four simulations of this type: two with the host
galaxy modeled by a polytropic distribution function and the shell
sampled with 20 and 100 fragments (runs $A1$ and $A2$,
respectively) and two with the galaxy described by the $f^{(\nu)}$
model and the same configurations for the shell (runs $A3$ and
$A4$). Runs $A1$ and $A2$ basically repeat runs $BS_1$ and $BS_2$
described in Paper I, but with the new improved code (Paper I made
use of the original van Albada (1982) code).

\begin{figure}
   \begin{center}
   \begin{tabular}{c}
     \includegraphics[angle=0,width=\textwidth]{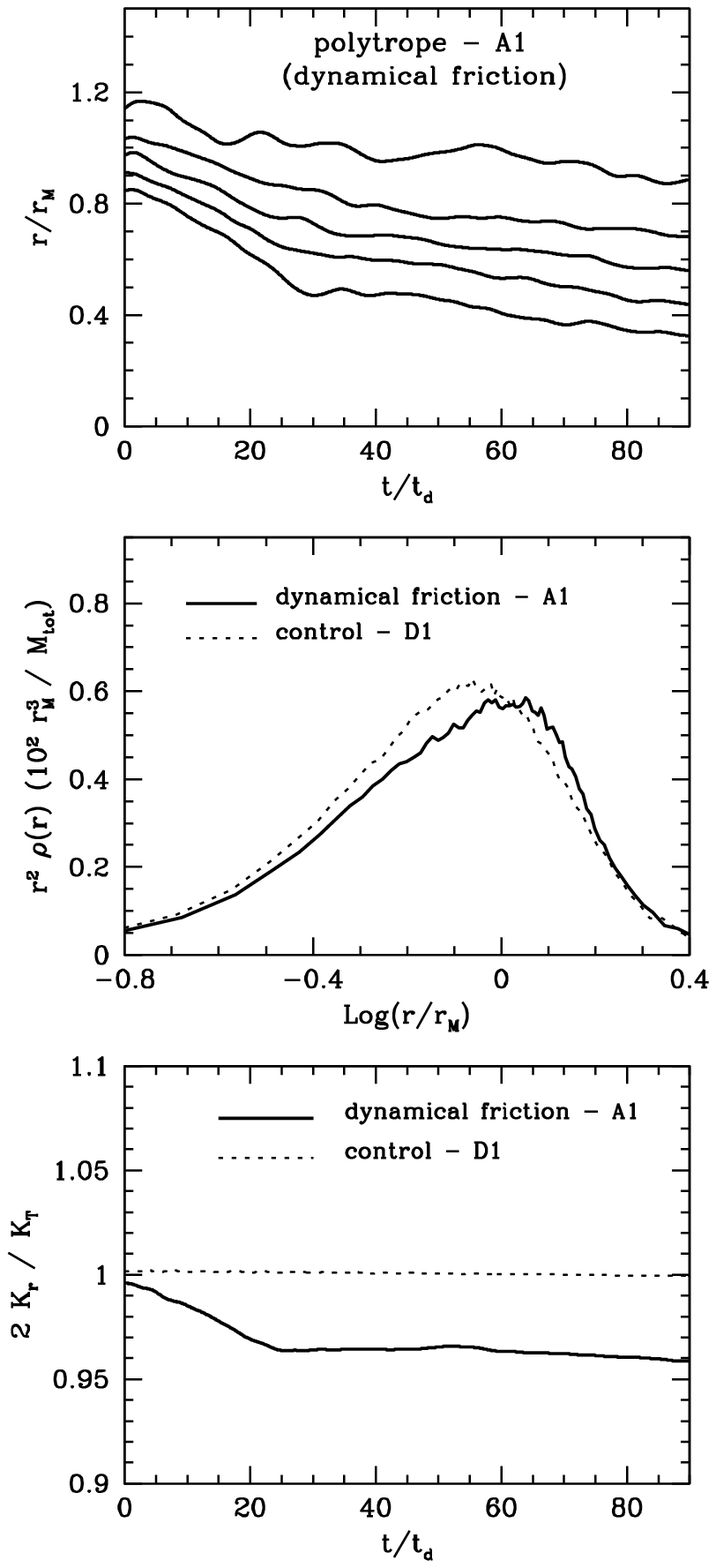}\\
     \end{tabular}
       \caption{Case A (dynamical friction): collisional shell of $N_{f}$ fragments
       in an $n = 3$ polytropic galaxy (run $A1$).
       The top panel describes the time evolution of the
       radii of the spheres containing 15, 35,
       55, 75, 95 \% of the total shell mass.
       The central panel compares the final galaxy
       density profile to that of the corresponding control case,
       which is very close to the initial galaxy density profile. The bottom
       panel illustrates the time evolution of the galaxy global
       anisotropy parameter ($2 K_r / K_T$).}
        \label{Fig01}
   \end{center}
   \end{figure}

In these runs, because of the dynamical friction experienced by
the individual fragments, the shell of fragments sinks in slowly,
toward the center of the galaxy, while becoming thicker in radius,
exchanging part of its energy with the galaxy (see top panel of
Fig.~\ref{Fig01} for run $A1$). For the same galaxy model, by
reducing the number of fragments (from 100 to 20) the fall of the
shell is faster, as expected because the fragments have larger
masses, and the thickening of the shell is less pronounced.

The host galaxy responds with an expansion that leads to a density
profile broader in the outer parts and shallower in the inner
regions. The degree of variation of the density profile from the
initial to the final configuration increases when the number of
shell fragments is decreased (from about 11\% of run $A1$ to about
2\% of run $A2$, for the polytropic model) and the size of the
effects changes with the galaxy model (from about 11\% of run $A1$
to about 25\% of run $A3$ for $N_{f}=20$). The initial and final
density profiles observed in run $A1$ are illustrated in the
central panel of Fig.~\ref{Fig01}; the plot makes use of the
volume-weighted density $r^2 \rho(r)$ so as to better bring out
the shift of mass associated with the observed evolution.

In addition, the host galaxy responds by changing its pressure
tensor in the tangential direction; the effect is more marked in
cases $A1$ and $A3$, which have a smaller number of fragments. For
the polytropic model, the final values of the global anisotropy
parameter $2K_r/K_T$ (twice the ratio of the total kinetic energy
in the radial direction to that in the tangential directions) are
0.96 and 0.98 respectively for runs $A1$ and $A2$ (see bottom
panel of Fig.~\ref{Fig01} for run $A1$), while the initial
configuration starts as isotropic (with $2K_r/K_T = 1$). In turn,
the $f^{(\nu)}$ model, which is characterized by a radially biased
pressure anisotropy to begin with, slowly evolves toward a more
isotropic configuration.

Because collisions are at the basis of the underlying physical
mechanism of dynamical friction, we expect to see their effects on
the particles of the galaxy that have taken part in the exchange
of energy and angular momentum with the fragments of the shell.
For the polytropic model, by comparing the first column of
Fig.~\ref{Fig03} with the first column of Fig.~\ref{Fig04}, which
represents the control case, one indeed sees that the
single-particle energy $E$, angular momentum $J$, and radial
action $I$ (of the galaxy simulation particles) are not conserved.
The percentage of particles with quantities conserved to a given
level increases with $N_{f}$. In the polytropic galaxy, about 66\%
and 45\% (respectively for $N_{f} = 100$ and $20$) of the
particles have $|(E_{fin} - E_{in})/E_{in}| < 0.08$, where
$E_{in}$ and $E_{fin}$ are the initial and final energy values, in
contrast with the value of 94\% observed in the collisionless
control case $D$. The corresponding values for $J$ and $I$ and for
the runs related to the $f^{(\nu)}$ model are shown in
Tab.~\ref{Tab03}. [The generally worse conservation of $J$ and $I$
with respect to that of $E$ depends on our definition of
conservation in terms of {\it relative} variations and on the fact
that a significant number of particles have values of $J$ and $I$
near zero. For example, in the control run $D1$, if we focus only
on particles with $J/J_{max}$ and $I/I_{max}$ greater than 0.25
(where $J_{max}$ and $I_{max}$ are the maximum values of $J$ and
$I$ taken by the galaxy particles), the percentage of conservation
increases respectively from 61 to 76 and from 38 to 65 (the
fraction of particles left out is 32\% and 70\%).]

   \begin{figure}
   \begin{center}
   \begin{tabular}{cc}
     \includegraphics[angle=0,width=\textwidth]{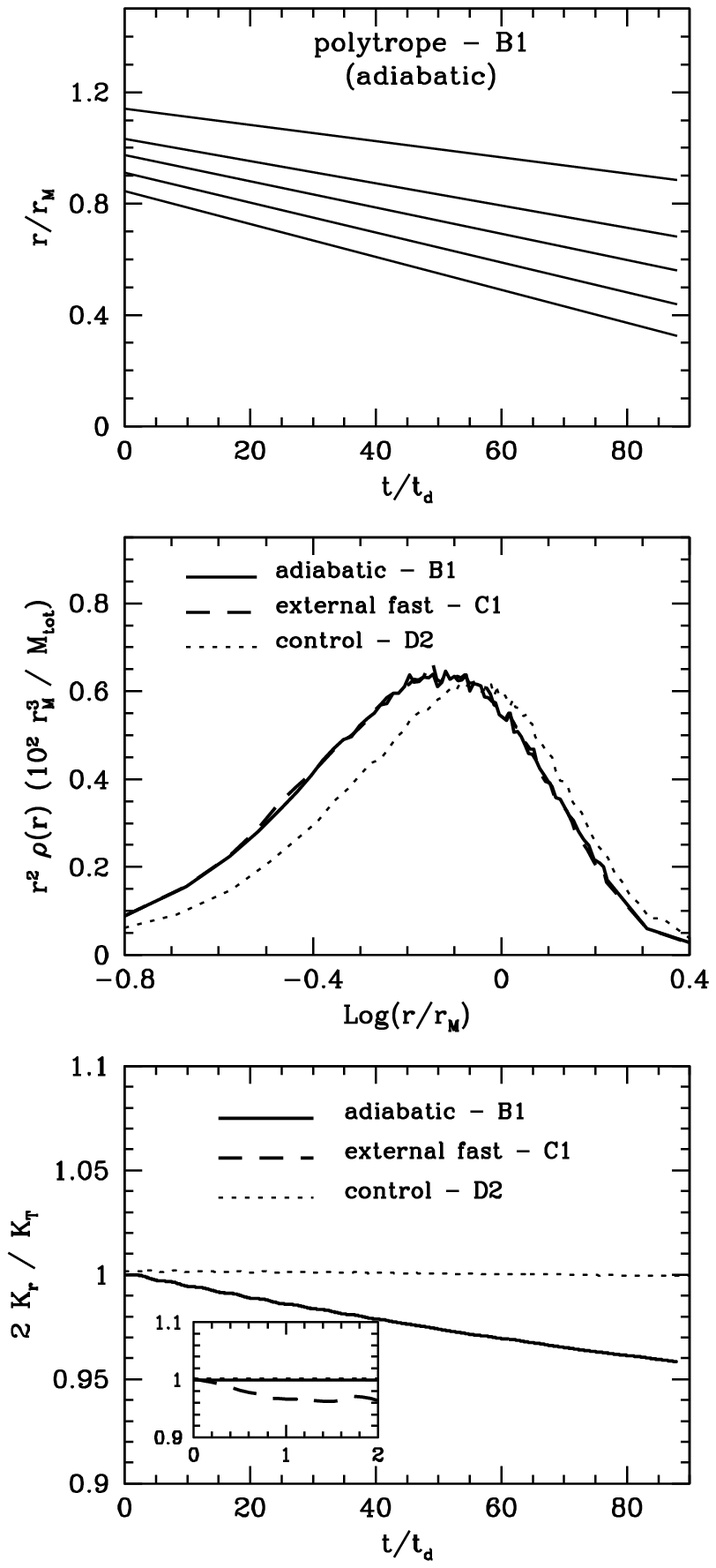}\\
     \end{tabular}
       \caption{Cases B (slow and adiabatic) and C (fast, non-adiabatic): evolution of an $n=3$
       polytropic galaxy in the presence of an external
       potential which mimics the fall of the shell observed in run $A1$.
       The three panels describe the same
       quantities as in Fig.~\ref{Fig01}. Note the qualitative difference with case
       $A$ in the evolution of the density profile. In the middle frame
       the $B$ and $C$ curves are basically identical. In the bottom frame the
       fast case $C1$ is illustrated only in the insert.}
         \label{Fig02}
   \end{center}
   \end{figure}

\subsection{Case B: Fall of a shell described by a slow external potential}

We have performed two simulations of this type: run $B1$ for the
polytropic galaxy model and run $B2$ for the $f^{(\nu)}$ model.

The construction of the external force mimicking the fall of a
shell is based on the results for the shell density distribution
with $N_{f} = 20$ observed in runs $A1$ and $A3$. The external
specific radial force to which a galaxy particle, at radius $r$
and at time $t$, is subject is given by $f_r(r,t) = -
GM_{shell}(r,t)/r^2$, where $M_{shell}(r,t) =
M_{shell}(r,t=0)(t_{fin}-t)/t_{fin} +
M_{shell}(r,t=t_{fin})(t/t_{fin})$ is the mass of the shell inside
the sphere of radius $r$ at time $t$. The values of
$M_{shell}(r,t=0)$ and $M_{shell}(r,t=t_{fin})$ are computed in
the runs $A1$ and $A3$ (for run $A1$, see top panel of
Fig.~\ref{Fig02}).

The observed evolution of the galaxy in the presence of the slow
change of the potential associated with the external shell is very
different from that in response to the direct interaction with the
fragments of the shell. This is a purely collisionless case in
which the galaxy energy changes slowly, much like in cases $A$,
but in the absence of dynamical friction.

The density distribution becomes more peaked in the inner regions
and shallower in the outer parts (see central panel of
Fig.~\ref{Fig02}). For a given galaxy model, the effect is found
to be stronger when a comparison is made with a case of dynamical
friction characterized by the lower value of $N_{f}$ ($N_{f} =
20$). At fixed value of $N_{f}$, the effect is stronger for the
$f^{(\nu)}$ model.

The total kinetic energies of the galaxy $K_r$ and $K_T$ (in the
radial and in the tangential directions) increase with time,
because of the energy change associated with the time-dependent
process, but at different rates, so that the galaxy global
pressure anisotropy evolves in the tangential direction. Much like
in runs of case $A$, the final configuration of the polytrope is
tangentially anisotropic (with a final global parameter $2K_r/K_T
= 0.96$; see bottom panel of Fig.~\ref{Fig02}). A similar trend is
found for the $f^{(\nu)}$ model.

At the microscopic level, as expected, the single-particle energy
is not conserved because the potential in which the particles move
is time-dependent. On the other hand, the angular momentum and the
radial action are conserved, within the numerical limits (see
second column of Fig.~\ref{Fig03} and Fig.~\ref{Fig04} and
Tab.~\ref{Tab03}). A similar behaviour is found for the
$f^{(\nu)}$ model.

\subsection{Case C: Fall of a shell described by a fast external potential}

As for case $B$, we have performed two simulations of this type:
run $C1$ for the polytropic galaxy model and run $C2$ for the
$f^{(\nu)}$ model.

In these simulations the external shell is moved in, from the
initial to the final state, as in case $B$, but at a much faster
speed, that is in only two dynamical times. This rapid variation
leads to a final configuration characterized by a density
distribution (see central panel of Fig.~\ref{Fig02}) and global
pressure anisotropy (see bottom panel of Fig.~\ref{Fig02}) similar
to those observed in case $B$. In particular, the curves labelled
$B1$ and $C1$ in the middle frame of Fig.~\ref{Fig02} overlap,
showing that the density response of the galaxy to the externally
imposed field mimicking the infall of a shell does not depend on
whether the process is actually slow or fast. A similar behaviour
is observed in the study of the $f^{(\nu)}$ model.

\begin{figure*}
   \centering
   \includegraphics[angle=-90,width=1.1\textwidth]{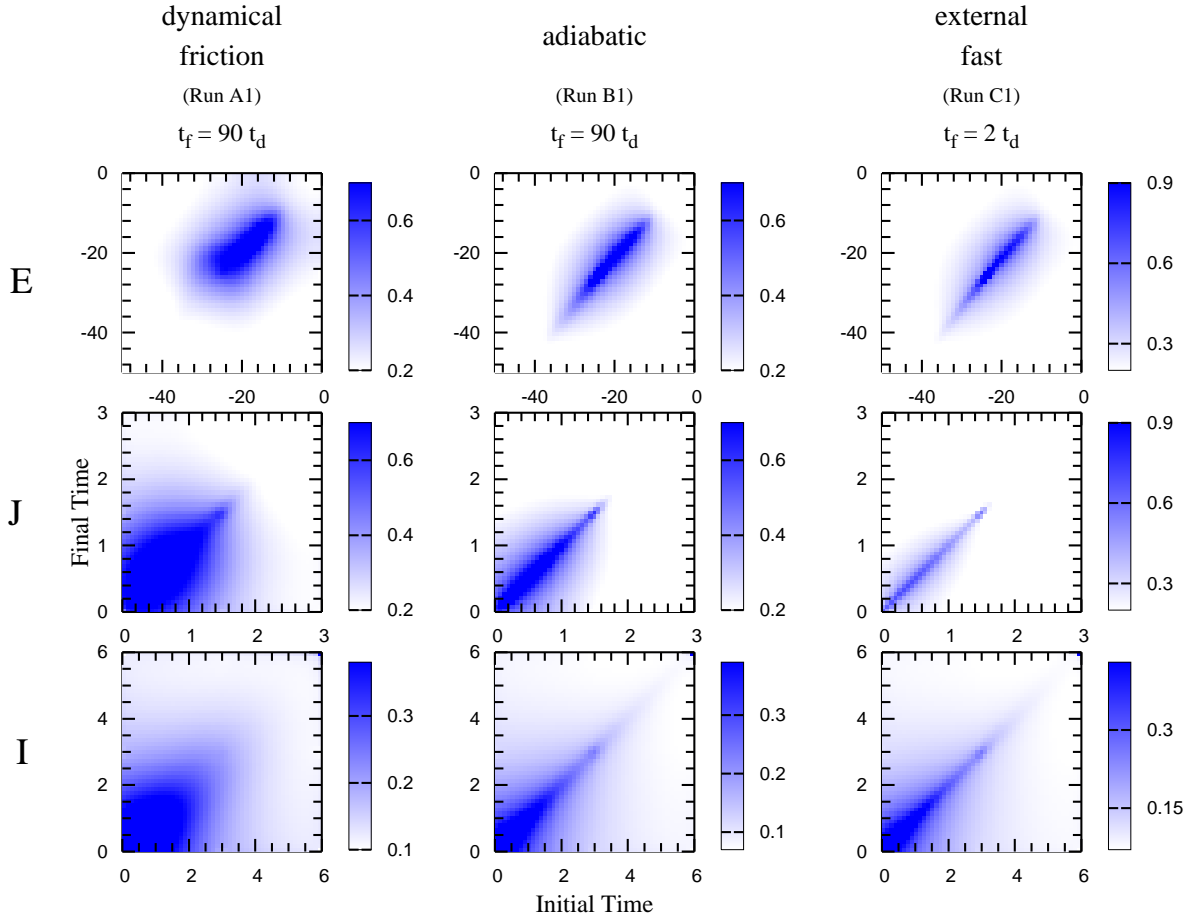}
   \caption{Correlation between single-particle quantities at an initial time
    ($t_{in} = 0.14~t_d$, x--axis) and at a final time ($t_{fin}$,
    y--axis): energy (top row), angular momentum (central row), and radial action
    (bottom row), expressed in code units (see Sect.~\ref{units}). 
    The first column displays the results for case
    $A$ (run $A1$; $t_{fin} = 90~t_d$), the second for case $B$ (run $B1$;
    $t_{fin} = 90~t_d$), the third for case $C$ (run $C1$; $t_{fin} = 2~t_d$);
    the corresponding control cases are displayed in
    Fig.~\ref{Fig04}. For each quantity $Q$, the grey scale used represents
    the number density of simulation particles in the plane
    $Q_0$--$Q$. By comparing each panel with the corresponding panel in Fig.~\ref{Fig04},
    the conservation properties listed in Tab.~\ref{Tab01} (and quantified in
    Tab.~\ref{Tab03}), are confirmed.}
              \label{Fig03}%
    \end{figure*}

At the microscopic level, for both galaxy models, the
single-particle energy and radial action are not conserved, while
the angular momentum is because of the spherical symmetry. The
numerical scatter observed in the relevant scatter plots is
relatively small because of the short duration of the simulation
(see third column of Fig.~\ref{Fig03} and Fig.~\ref{Fig04} and
Tab.~\ref{Tab03}).

\begin{table}
\begin{minipage}[t!]{\columnwidth}
\caption{Conservation of single-particle quantities.
For the various runs of cases $A$ and $B$ the Table lists the
percentage of galaxy particles with $|(Q_{fin} - Q_{in})/Q_{in}| <
0.08$ for the quantities $Q=E$, $J$, and $I$. For case $C$ the
``conservation level" is lowered to 5\%, because the scatter is
smaller due to the short duration of the simulations. For each
case ($A$, $B$, and $C$) the numbers are listed close to those of
the corresponding control case $D$ (cf. the expectations of
Tab.~\ref{Tab01}). The worse conservation of $J$ and $I$ with
respect to that of $E$ depends on the fact that a significant
number of particles have initial values of $J$ and $I$ near zero.}               
\label{Tab03}      
\centering                          
\renewcommand{\footnoterule}{}  
\begin{tabular}{c | c c c | c c | c c}        
\hline\hline                 

 \textbf{Case} & \textbf{D} & \multicolumn{2}{c|} {\textbf{A}} &  \textbf{D} & \textbf{B} & \textbf{D} & \textbf{C} \\
 $t_{in}$ | $t_{fin}$ & \multicolumn{3}{c|} {0.14 | 90 $t_d$} & \multicolumn{2}{c|} {0.14 | 90 $t_d$} &\multicolumn{2}{c} {0.14 | 2 $t_d$} \\
\hline\hline
 \multicolumn{8}{c} {\textbf{Polytrope  $(n=3)$}} \\
\hline
 Runs & D1 & A1 & A2 & D2 & B1 & D2 & C1 \\
\hline
 \textbf{$E$}  & 93 & 38 & 62 & 98 & 58 & 100 & 28 \\
 \textbf{$J$}  & 61 & 23 & 34 & 68 & 67 & 98 & 98 \\
 \textbf{$I$}  & 38 & 15 & 24 & 42 & 41 & 83 & 55 \\
\hline                                
  \multicolumn{8}{c} {\textbf{$f^{(\nu)}$}       \textbf{$(\nu, \Psi) = (3/4 , 5)$}} \\
\hline

 Runs & D3 & A3 & A4 & D4 & B2 & D4 & C2 \\
\hline
 \textbf{$E$}  & 89 & 30 & 57 & 91 & 42 & 100 & 29 \\
 \textbf{$J$}  & 55 & 20 & 32 & 60 & 60 & 89 & 89 \\
 \textbf{$I$}  & 46 & 18 & 31 & 48 & 48 & 83 & 54 \\
\hline
\end{tabular}
\end{minipage}
\end{table}

\subsection{Case D: Control runs with a collisionless and with a static external shell}

We have performed four simulations of this type: runs $D1$ and
$D2$ for the polytropic galaxy model and runs $D3$ and $D4$ for
the $f^{(\nu)}$ model.

These simulations have been performed to test the stability of the
initial conditions of the system considered in cases $A$, $B$, and
$C$. In runs $D1$ and $D3$, designed to test runs $A1$--$A4$, the
shell is sampled with $25000$ particles, so that each shell
simulation particle has the same mass as each galaxy simulation
particle; the shell particles are then treated as collisionless,
interacting with one another and with the the particles of the
galaxy via the mean field. In runs $D2$ and $D4$, designed to test
runs $B1$, $B2$, $C1$, and $C2$, the shell is treated as an
external field, much like in cases $B$ and $C$, but kept static,
in its initial configuration $M_{shell}(r,t=0)$.

   \begin{figure*}
   \centering
   \includegraphics[angle=-90,width=1.1\textwidth]{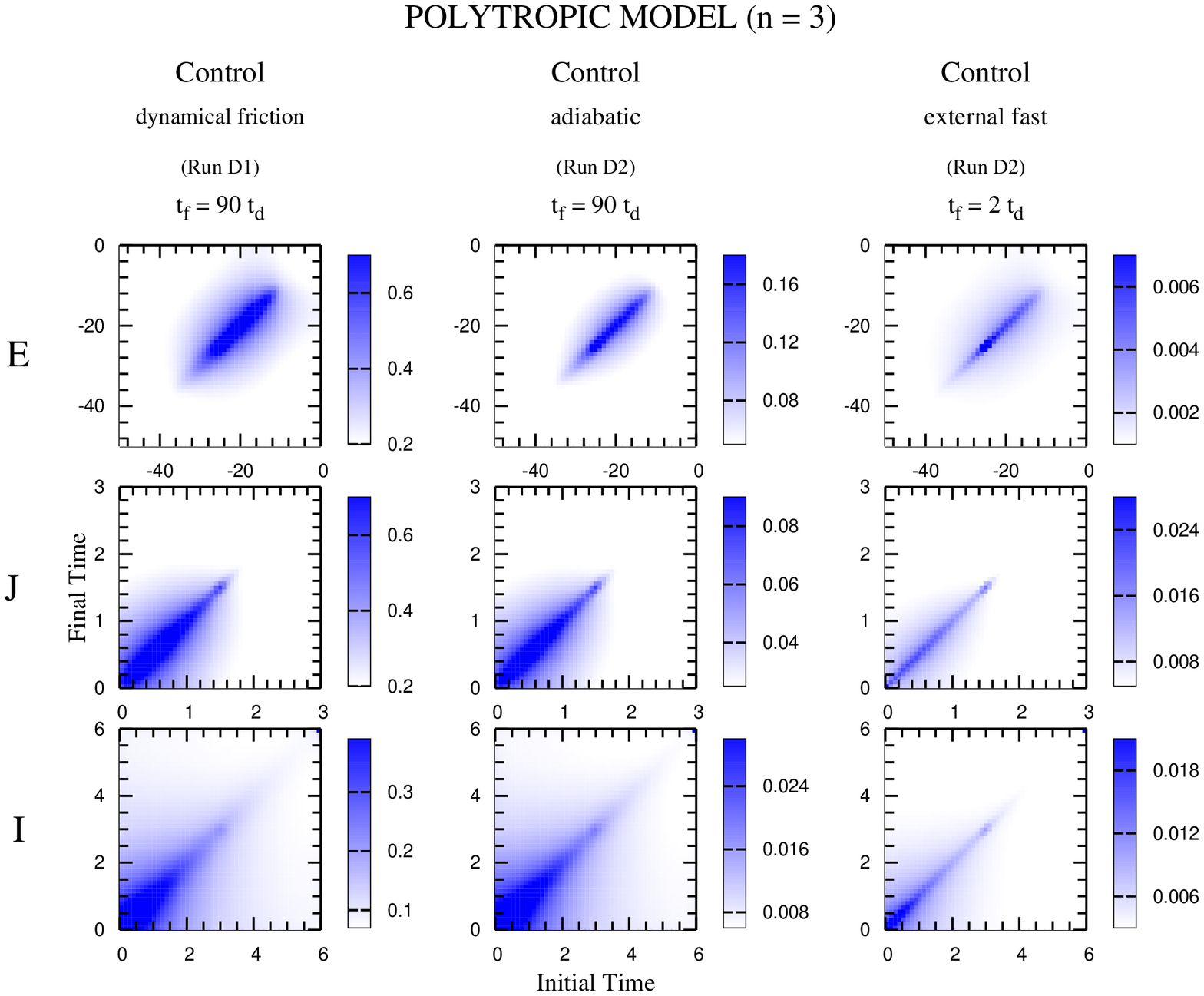}
   \caption{Correlations observed in the control runs $D1$ and $D2$,
   defining the accuracy of the code. The layout and the notation are similar to those
   of Fig.~\ref{Fig03}. The apparent thinness of the scatter plots
   in the third column of this figure and in
   Fig.~\ref{Fig03} are due to the shortness of the run presented ($t_{fin} =
   2~t_d$).
   }
              \label{Fig04}%
    \end{figure*}

The initial ``equilibrium" configuration of the host galaxy in the
presence of the collisionless or static shell is checked to stay
basically unchanged in the course of the simulations. For both
galaxy models, in run $D1$ and $D3$, the radii of the spheres
containing 15, 25, 35, 45, 55, 65, 75, 85, 95 \% of the total mass
of the shell remain constant in time to within 1\%, except for
that containing 5\% that remains constant only to 4\%. A similar
diagnostics for the galaxy simulation particles shows a
conservation better than 0.5\%; in the innermost regions ($r < 0.1
r_M$), because of the relatively small number of particles that
are involved, the density profiles are found to be steady to
within 2\%. The changes in the total kinetic energy of the galaxy
in the radial and tangential directions are found to be less than
0.3\%.

In runs $D2$ and $D4$ the overall stability is checked to be even
better than for runs $D1$ and $D3$: the mass radii remain constant
to within 0.05\% (except for the 5\% mass radius for which the
stability is to 0.2\%), the density profile remains unchanged to
within 1\%, and the total kinetic energies are stable to less than
0.1\%.

Because of these control cases are designed to be spherically
symmetric and time-independent, the single particle quantities
(energy $E$, angular momentum $J$, and radial action $I$) should
be conserved. The level of conservation observed over the entire
period of the simulation, thus illustrating the accuracy of our
code, is shown in Fig.~\ref{Fig04} and in Tab.~\ref{Tab03}.

For completeness, we have also run one additional control case,
$D5$, of a pure polytropic model, with no shell, to test whether
some of the scatter present in the control cases $D$ might be
associated with the introduction of the shell itself. The relevant
conservation levels over the time interval $0.14~t_d$ - $90~t_d$,
corresponding to the $D2$ column at the center of
Tab.~\ref{Tab03}, are 97 (for $E$), 66 (for $J$), and 41 (for
$I$), practically identical to the corresponding values for run
$D2$.

\subsection{One qualitative difference in behavior between the polytropic
and the $f^{(\nu)}$ model}

We have noted that, as a result of the slow evolution associated
with the process of dynamical friction (see middle frame of
Fig.~\ref{Fig01}), the density distribution of the galaxy is
softened, with the mass being basically transported outwards, in
contrast with the trend associated with the corresponding
adiabatic case (see middle frame of Fig.~\ref{Fig02}). This
general property is found in both the polytropic and the
$f^{(\nu)}$ models that we have investigated.

\begin{figure}
\begin{center}
\begin{tabular}{cc}
\includegraphics[angle=0,width=\textwidth]{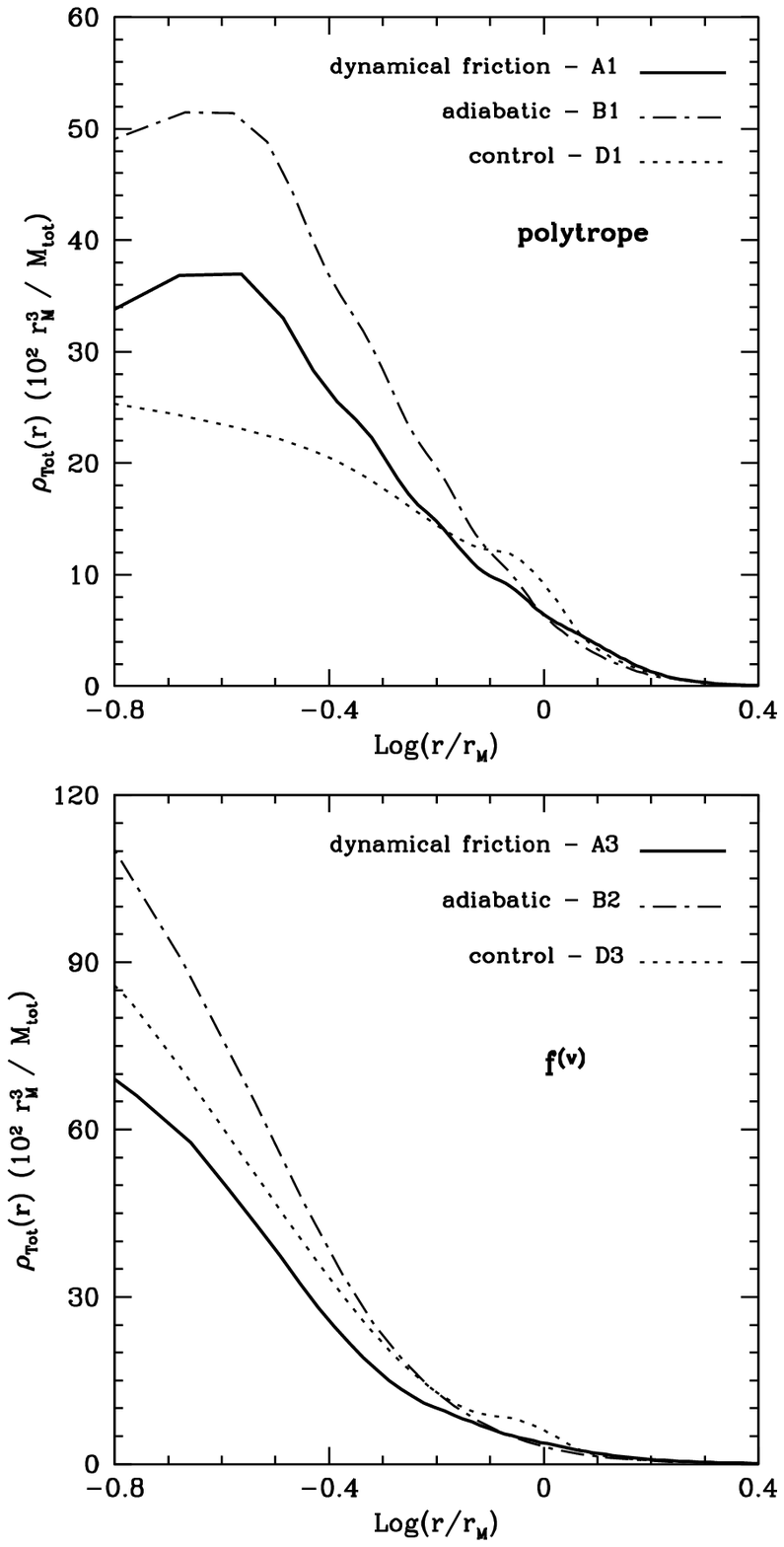}\\
\end{tabular}
\caption{Comparison of final total density profiles in different
evolution processes (solid lines, dynamical friction; dotted
lines, control cases; dash-dotted lines, adiabatic evolution) for
the polytropic galaxy model (upper frame) and for the $f^{(\nu)}$
model (lower frame). The control cases, which are basically
time-independent, thus illustrate the adopted initial conditions.}
\label{DensTot}
\end{center}
\end{figure}

If we focus instead on the evolution of the {\it total} density
profile (galaxy + shell), we note a curious qualitative difference
in the behavior of the two models. In fact, while for the
polytropic model (characterized initially by a broad core) the
evolution induced by dynamical friction leads to a higher
concentration in the {\it total} density profile, for the
$f^{(\nu)}$ model (which starts with higher concentration) the
softening and broadening of the density profile occurs not only
for the galaxy density distribution but also for the {\it total}
density profile (see Fig.~\ref{DensTot}). In all the cases
studied, the adiabatic evolution is associated with the most
concentrated final density profiles.

These considerations are especially relevant when one imagines
astrophysical applications based on the study of rotation curves
in the innermost regions of galaxies. We do not have a simple
physical argument to explain the observed difference in behavior,
which was noted as a side-result in this project dedicated to
studying the non-adiabatic character of dynamical friction. We are
planning to return to this issue in a separate paper that will be
devoted to comparing the process of dynamical friction in models
characterized by a broad core to the process occurring in
concentrated models.

\section{Conclusions and perspectives}

In this paper we have studied in detail the slow evolution of
quasi-collisionless self--gravitating systems by comparing the
situation when such evolution is due to effects of dynamical
friction to a situation that is strictly adiabatic but otherwise
similar. The two types of evolution have been studied by specially
designed numerical experiments and have been compared both at the
macroscopic and at the microscopic level.

At the microscopic level we have checked that the relevant
single-particle quantities, such as energy $E$, angular momentum
$J$, and radial action $I$, are indeed conserved or not conserved
as expected from first principles. These studies have been
quantified by means of suitable ``scatter plots" which 
bring out interesting structural properties of the
collisionless Boltzmann (Vlasov) equation, when suitably expressed
in terms of evolution variables (see Appendix).

At the macroscopic level, we find a qualitatively different
behaviour in the evolution of the density distribution of the
stellar system under investigation. In particular, in the
(non-adiabatic) case of slow evolution resulting from dynamical
friction, the density profile becomes ``softer" in the course of
evolution, in contrast with the overall steepening and contraction
noted in the adiabatic case. In turn, the overall evolution of
pressure anisotropy turns out to be similar in the two cases.

In this study we have considered the evolution induced on the host
galaxy by a slow infall of matter distributed in a relatively thin
shell. This paper is part of a longer-term project, currently in
progress, where we aim at studying the more general problem in
which the infalling matter has a full three-dimensional
distribution. By comparing such more general behavior with the
behavior that may be anticipated by superposing the effects of
adjacent thin shells of matter (now described quantitatively in
this paper), we will be able to assess to what extent the general
problem of infall can be reduced to a {\it local} analysis or else
whether full global simulations are generally required, with a
behavior that may depend on a case by case basis. It would be
important to identify general trends in such general process, even
if those trends will be different from those suggested by the
picture of adiabatic infall.

The astrophysical applications of this long-term project are
basically directed on two goals. Specifically in the galactic
context, we wish to develop a quantitative picture of the effects
on a stellar system (such as a bulge or an elliptical galaxy)
associated with the slow infall of a system of globular clusters;
similar effects may operate in determining a more realistic
picture of galaxy formation, in which the capture of a large
number of satellites (minor mergers) may accompany or follow the
process of incomplete violent relaxation (known to occur in purely
collisionless systems). In the more general cosmological context,
we wish to assess whether dynamical friction can change
significantly the scenario of cusp formation, so prominently
present in cosmological simulations. For addressing properly all
these issues we will have to extend the calculations so as to
allow for the evaporation and possible tidal disruption of the
satellites that are dragged in by dynamical friction.

\begin{acknowledgements}

We would like to thank Michele Trenti for providing us with his
improved code for the simulations and Luca Ciotti, Michele Trenti,
and Tjeerd van Albada for a number of useful suggestions. We are
also glad to thank Alessandro Romeo for his constructive
criticism. Part of this work was supported by the Italian MIUR.

\end{acknowledgements}

\appendix
\section{}

We start by examining the origin of distortions and dispersion in
scatter plots for the ideal case in which the simulation particles
follow rigorously and exactly the characteristics of the
collisionless Boltzmann equation. To set a quantitative framework
for the corresponding effects associated with the numerical noise
is far more difficult and beyond the scope of the present paper.
The main idea that we follow below consists in reading the scatter
plots as suitable cuts (or projections) of a distribution function
expressed in terms of evolution variables.

A formal definition of the scatter plots mentioned above can be
given in terms of the distribution function $f=f({\bf x},{\bf
p},t)$ of a stellar system (i.e., more precisely of its  mass
density distribution function). Here  ${\bf x},{\bf p}$ are the
$6$ position and momentum coordinates in phase space and $f({\bf
x},{\bf p},t)~ d{\bf x}~ d{\bf p}$ is the mass inside the phase
volume $d{\bf x}~ d{\bf p}$. The distribution $f$ obeys the
collisionless Boltzmann equation

\begin{equation} \label{V}
\frac{\partial f}{\partial t} +  \{f, H\}_{x,p} = 0,
\end{equation}

\noindent where $H=H(x,p,t)$ is the single-star Hamiltonian that
includes the effect of external forces such as those arising from
the presence of the falling satellites and $\{~, ~\}_{x,p} $ are
the standard Poisson brackets (referred to the variables ${\bf x}$
and ${\bf p}$).  As is well known, Eq. (\ref{V}) can be rewritten
in the Lagrangian form

\begin{equation} \label{L}
\frac {D f}{D t} = 0,
\end{equation}

\noindent which expresses the fact that the  distribution function
remains constant as it evolves in time along the characteristics
of the Hamiltonian $H$. In turn, the general solution of Eq.
(\ref{L}) is $f({\bf x},{\bf p},t) = F({\bf x}_0,{\bf p}_0)$ where
${\bf x}_0,{\bf p}_0$  are the phase-space coordinates at $t =0$
of the characteristics that at time $t$ reach ${\bf x},{\bf p}$
and $F$ denotes the assigned initial condition for the
(positive-definite) distribution function at $t =0$, i.e. $f({\bf
x}_0,{\bf p}_0,t=0) = F({\bf x}_0,{\bf p}_0)$. Note that $d{\bf
x}~ d{\bf p} = d{\bf x}_0~ d{\bf p}_0$, i.e., the Jacobian of the
coordinate transformation in phase space from ${\bf x},{\bf p}$ to
${\bf x}_0,{\bf p}_0$ is equal to unity (corresponding to the
invariance properties of one of the so-called Poincar\'{e}
integrals). Clearly in the ${\bf x}_0,{\bf p}_0$ coordinates the
Vlasov equation reads

\begin{equation} \label{Vin}
\frac{\partial F}{\partial t} = 0,
\end{equation}

\noindent where the partial time derivative is taken at constant
${\bf x}_0,{\bf p}_0$.

Suppose we are interested in studying the non-conservation of a
given quantity $Q$ (which might be the single-star energy or
radial action). Then we may introduce a different coordinate
system in phase space where instead of ${\bf x}_0,{\bf p}_0$ we
introduce the pair $Q_0,~ Q$, where $Q = Q(t)\equiv Q({\bf
x}(t),{\bf p}(t),t)$ is the appropriate function of the phase
space coordinates ${\bf x},{\bf p}$ and of time and $Q_0 =
Q(t=0)$, supplemented by four initial coordinates, which we
symbolically denote by $\alpha_j$, with $j = 1,..,4$. If $Q$ is
not conserved, we expect that for $t \not= 0$  the Jacobian $
\mathcal{J} = \mathcal{J} (Q_0,Q,\alpha_j; {\bf x}_0,{\bf p}_0)$
be different from zero; but at $t=0$ we have $Q = Q_0$, so that
the coordinate transformation becomes singular and $\mathcal{J} =
0$. Clearly, if $Q$ were a conserved quantity the above coordinate
transformation would be singular at all times; a similar situation
may occur in the presence of an adiabatic invariant, when a
quantity $Q$, such as the single-particle energy, is uniquely
determined once $Q_0,\alpha_j$, and $t$ are given.  In addition,
the Jacobian $\mathcal{J}$ may vanish at special values of $t$,
for special values of $Q_0,Q$ and $\alpha_j$, when a ``caustic"
type behaviour occurs.

In terms of the coordinates $Q_0,Q,\alpha_j$ the Vlasov equation
for $\hat{F} = \hat{F}(Q_0,Q ,\alpha_j, t) = F ({\bf x}_0,{\bf
p}_0)$ reads

\begin{equation} \label{Vev}
\frac{\partial \hat{F}}{\partial t} + \frac{D Q(t)}{D t}~
\frac{\partial \hat{F}}{\partial Q} = 0,
\end{equation}

\noindent where ${D Q(t)}/{D t}$ is to be expressed as a function
of $Q_0,Q,\alpha_j$, and $t$. Here the partial time-derivative is
taken at constant $Q_0,Q ,\alpha_j$. In terms of $\hat{F}$, the
mass density $f({\bf x},{\bf p},t)~ d{\bf x}~ d{\bf p}$  becomes

\begin{equation} \label{dens}
f({\bf x},{\bf p},t)~ d{\bf x}~ d{\bf p} ~=~ \frac{\hat{F}(Q_0,Q
,\alpha_j,t)}{\mathcal{J}(Q_0,Q ,\alpha_j)} ~ d Q_0 ~d Q
~d\alpha_j.
\end{equation}

\noindent The general solution of Eq.~(\ref{Vev}) can be written
as

\begin{equation} \label{solut}
{\hat{F}(Q_0,Q ,\alpha_j,t)} = {\cal F}\big(Q_0, \alpha_j,
K_{\alpha_j}(Q_0, Q,t)\big) ,
\end{equation}

\noindent where the characteristic $K_{\alpha_j}(Q_0, Q,t)$ is a
solution of Eq.~(\ref{Vev}) at fixed $\alpha_j$.

An explicit solution of the above equations is given at the end of this
Appendix for the special case of a one-dimensional motion in a
time dependent but spatially uniform force. Here we consider  two
paradigmatic examples. Let us first consider the case of

\begin{equation} \label{E1}
\frac{D Q(t)}{Dt} = \frac{(Q - Q_0)}{t},
\end{equation}

\noindent which is expected to hold as typical behavior in the
initial stages of the evolution, i.e., for $t\to 0$. At fixed
$Q_0$, the characteristic is then given by

\begin{equation} \label{E2}
K_{\alpha_j}(Q_0, Q,t) = \frac{(Q - Q_0)}{t},
\end{equation}

\noindent so that the general solution for $\hat{F}$ should be
written as a function ${\cal F}$ of the variables $Q_0$,
$\alpha_j$, and $K_{\alpha_j}$:

\begin{equation} \label{E3}
\hat{F}(Q_0, Q, \alpha_j, t)=  {\cal F}\big(Q_0, \alpha_j,  (Q -
Q_0)/t\big).
\end{equation}

\noindent Equation (\ref{E3}) provides the generic form of the
distribution function in evolution variables for early times such
that $(Q-Q_0)/Q_0\ll 1$ and shows that if the initial distribution
function is only a function of $Q_0$ and of $\alpha_j$, it remains
independent of $Q$ at all times (over the allowed domain of $Q$ at
time $t$).

In the second example we take

\begin{equation}
\frac {D Q(t)}{Dt} =  Q \frac{D
\ln{(\omega(Q_0,\alpha_j,t))}}{Dt},
\end{equation}

\noindent we have

\begin{equation} \label{E4}
K_{\alpha_j}(Q_0, Q,t) =   Q/\omega
\end{equation}

\noindent where $\omega= \omega(Q_0,\alpha_j,t)$ is an effective
``frequency" and the allowed range of $Q$ at fixed $Q_0,\alpha_j$
is restricted to the single value $Q = \omega Q_0/\omega_0$, with
$\omega_0 = \omega(Q_0,\alpha_j,t=0)$. Therefore, in this case we
find $Q -Q_0 = Q_0 (\omega/\omega_0- 1)$, which compared with
Eq.~(\ref{E3}), illustrates the fact that in the presence of an
adiabatic invariant the transformation from ${\bf x},{\bf p},t$ to
$Q_0,Q,\alpha_j,t$ is singular at all times. In turn, finding a
high concentration of points at all times along a curve in a
scatter plot would suggest that the transformation be close to
singular, which provides us with a global test of adiabatic
invariance.

The scatter plots mentioned in Subsection \ref{diag} and used in
Sect.~\ref{sim} thus correspond to the mass density $D(Q_0, Q,t)$
in the $(Q_0,Q)$ plane obtained by integrating at time $t$ the
mass density in phase space  over the coordinates $\alpha_j$

\begin{equation} \label{scat}
D(Q_0, Q,t)  = \int   d\alpha_j ~\frac {{\cal F}\big(Q_0,
\alpha_j, K_{\alpha_j}(Q_0, Q,t)\big)}{ \mathcal{J}(Q_0,Q
,\alpha_j)}.
\end{equation}

\noindent For $t\to 0$ we have

\begin{equation} \label{scat2}
D(Q_0, Q,t)  =  \frac{{\cal D}(Q_0, (Q- Q_0)/t)}{t}.
\end{equation}

The framework that we have developed suggests a number of tests
and of scatter plots based on different variables, especially in
the limit $t\to 0$. We postpone these investigations to a separate
paper, because they touch on issues that are not of immediate
astrophysical interest, beyond the main goals of this article.

To conclude, we exemplify the theoretical framework just outlined
by considering one simple time-dependent system for which the
characteristics can be easily computed.

We consider the one-dimensional motion of a star in a
homogeneous time dependent gravitational field described by the
Hamiltonian

\begin{equation} \label{A1}
H(x,p,t) = p^2/2  - 2 xt.
\end{equation}

\noindent The characteristics are easily computed and are given by
\begin{eqnarray} \label{A2}
p &=& p_0 +  t^2, \\x &=& x_0 + p_0 t + t^3/3,\nonumber
\end{eqnarray}

\noindent so that

\begin{equation} \label{A3}
H(x_0,p_0,t) = p_0^2/2 -2x_0t -p_0 t^2 -t^4/6.
\end{equation}

\noindent Note that $H_0 = H_0(x_0,p_0, t=0) = p_0^2/2$.

We now focus on the non-conserved quantity $H$ (which was called
$Q$ above) and consider the transformation
$(x_0,p_0) \rightarrow (H_0, H)$, which is easily recovered from:

\begin{eqnarray} \label{A4}
p_0 &=& (2 H_0)^{1/2},\nonumber\\
x_0 &=&  (H_0 - H)/(2t) - (2H_0)^{1/2} t/2 - t^3/12.
\end{eqnarray}

\noindent Here the square roots are properly defined so as to keep
track of the sign of $p_0$. From Eq.~(\ref{A3}) we obtain

\begin{equation} \label{A5}
\frac{D H}{Dt} =  - 2 x_0 - 2 p_0 t - \frac{2 t^3}{3} = - 2 x =
\frac{\partial H}{\partial t},
\end{equation}

\noindent which can be rewritten as

\begin{equation} \label{A6}
\frac{D H}{Dt} = \frac{H - H_0}{t} - (2 H_0)^{1/2} t -
\frac{t^3}{2},
\end{equation}

\noindent and reduces to $D H/Dt \sim (H - H_0)/t$ for $t\to 0$.
The characteristic function  $K(H_0, H,t)$ defined by
Eq.~(\ref{Vev}) is given by

\begin{eqnarray} \label{A7}
  K(H_0, H,t) &=&  (H_0 - H)/(2t) - (2 H_0)^{1/2} t/2 - t^3/12 \nonumber\\
  &=& x_0(H_0, H,t)~~~
\end{eqnarray}

\noindent as can be easily verified by substitution and is {\it a
priori} obvious from the fact that, at constant $H_0$, the
characteristics are labelled by the value of $x_0$.

The Jacobian $\mathcal{J}(H_0,H)$ is given by

\begin{equation} \label{A8}
 \mathcal{J}(H_0, H) =  2 |p_0 t| = 2 (2 H_0 t^2)^{1/2}.
\end{equation}

\noindent Thus from the distribution function $F(x_0,p_0)$, in the
$H,H_0, t$ variables we have

\begin{equation}\label{A9}
\begin{array}{lr}
F(x_0,p_0) ~d x_0 d p_0 =&\nonumber\\
& \frac{{\cal F}(K(H_0, H,t),
(2 H_0)^{1/2})}{2(2 H_0 t^2)^{1/2}}~ d H_0 d H,
\end{array}
\end{equation}

\noindent with $K(H_0, H,t)$ defined by Eq.~(\ref{A7}). Note that,
if at $t=0$ the distribution function is homogeneous in space it
is constant in $H$ at all times.  This is best realized by
considering a distribution that is Gaussian both in $p_0$ and in
$x_0$, for which we obtain in $H,H_0,t$

\begin{eqnarray}
\exp{(-p_0^2/2  - x_0^2/L^2)}\frac{ d x_0 d p_0}{L}= \nonumber\\
\exp{\{-[(H - H_0)/(2t) + (2 H_0)^{1/2} t/2 +
t^3/12]^2/L^2\}}\times \nonumber \\ \exp{(-H_0)} \frac{d H_0 d
H}{2L(2 H_0 t^2)^{1/2}},~~~~~
\end{eqnarray}

\noindent which becomes independent of $H$ in the limit $L\to
\infty$. For finite $L$, in the limit $t\to \infty$ the
distribution function is centered around $H = - t^4/6$ and has a
width that increases much slower, as $t^{1/2}$, i.e., the
distribution tends to become ``monochromatic''.

Two different examples, the case of a parametric linear oscillator
and the case of a free star in a small-amplitude wave-like
potential, have also been solved explicitly in this framework,
demonstrating how adiabatic invariance, the formation of caustics,
and the effects of resonances may be recognized by means of
studies of the Vlasov equation in evolution variables.

%

\end{document}